\documentclass[prl,aps,floats,twocolumn,superscriptaddress,tightenlines,notitlepage,preprintnumbers,showpacs,amsmath,amssymb]{revtex4}

\usepackage{epsfig}
\usepackage{amsmath}

\begin{document}
%

%
%

%
%

\title{ Connecting black holes and black strings }

\author{Hideaki Kudoh}
\email{kudoh@utap.phys.s.u-tokyo.ac.jp}
\affiliation{
Department of Physics, The University of Tokyo, Bunkyo-ku, 113-0033, Japan   
}

\author{Toby Wiseman}
\email{twiseman@fas.harvard.edu}
\affiliation{Jefferson Physical Laboratory, Harvard University, Cambridge MA 02138, USA}

\date{September 2004}

\preprint{hep-th/0409111, UTAP-495 } 


%
\begin{abstract}
%

Static vacuum spacetimes with one compact dimension include black
holes with localised horizons but also uniform and non-uniform black
strings where the horizon wraps over the compact dimension. We present
new numerical solutions for these localised black holes in 5 and
6-dimensions. Combined with previous 6-d non-uniform string results,
these provide evidence that the black hole and non-uniform string
branches join at a topology changing solution.

%
\end{abstract}
%

\maketitle

\section{Introduction and Summary}
\label{sec:intro}

The task of this letter is to resolve the structure of static vacuum
solutions of Kaluza-Klein theory, namely pure gravity
compactified on a circle. Let us firstly motivate our interests in
this problem.

Many scenarios in string theory have extra dimensions large enough
that they may be described geometrically \cite{ADD,AADD}. In such
models it is important to understand the behaviour and types of black
hole solutions, and in particular whether there are potentially new
signals from this physics \cite{Kol:2002hf}.

The problem is also interesting as it is connected by holography to
the phase structure of large $N_c$ super Yang-Mills theory,
compactified on a circle, at strong t'Hooft coupling
\cite{Itzhaki:1998dd,Susskind:1998vy,Martinec:1998ja,Li:1998jy,dual,Harmark:2004ws}. Since
strong coupling results are scarce for this field theory, gravity
provides the only window into this regime.  Furthermore, the same
phase structure predicted by gravity at strong coupling appears to
persist to weak coupling \cite{dual}.

Lastly, this letter provides evidence that for Kaluza-Klein theory the
3 branches of solutions - the localised black holes, uniform and
non-uniform strings - are connected in an elegant way, initially
conjectured by Kol who postulated the problem is controlled by one
relevant order parameter \cite{Kol1}. Learning more about this
Morse-theory inspired approach may shed light on the new and exotic
phenomena found in higher dimensions
\cite{Emparan_Reall2,Sorkin:2004qq,Kol:2004pn}.

For small masses localised black holes (BH) should look like
Schwarzschild solutions. Increasing their mass they deform as they
feel the compact circle\cite{Harmark:2003yz,Gorbonos:2004uc}. The key
question is then whether there is a maximum size for these solutions
beyond which they no longer `fit'.

In $d$ spacetime dimensions, with $d \ge 5$, uniform string (US)
solutions exist. These are direct products of $(d-1)$ dimensional
Schwarzschild with the circle, and are the only uncharged solution with horizon that is asymptotically $R^{1,d-2} \times S^1$ which has a simple analytic form (except for bubble spacetimes which we do not consider
here \cite{bubbles}).

The non-uniform strings (NUS) were discovered when Gregory and
Laflamme showed that for a given circle size, uniform strings below a
critical mass are linearly unstable
\cite{Gregory_Laflamme1,Choptuik:2003qd}. At the critical point there
is a static mode breaking translation invariance on the
circle. Motivated by dynamical considerations \cite{Horowitz_Maeda1},
Gubser showed this mode remains static to all orders in perturbation
theory \cite{Gubser}. 

Kol conjectured \cite{Kol1} that the `waist' of the non-uniform string
would shrink to nothing, locally forming a cone geometry, which
connected to the black hole branch by resolving the cone to change the
horizon topology (see also \cite{Harmark_Obers}).  Elliptic numerical
methods were used to construct the non-uniform strings in 6-d
\cite{Wiseman1,Wiseman2} and the cone geometry was seen to emerge for
the most non-uniform solutions \cite{Kol_Wiseman}. These methods have
recently been applied to construct the black hole solutions in 5-d and
6-d, and thus in principle we can test Kol's conjecture from the
`other side' \cite{Kudoh_Wiseman,Sorkin:2003ka}.

Here we present new numerical solutions for the 5-d and 6-d localised
black holes which significantly improve on the previous works
\cite{Sorkin:2003ka,Kudoh_Wiseman} (see footnote \footnote{The black
hole calculations were improved over our previous computation
\cite{Kudoh_Wiseman} largely by increasing the numerical
resolution. In addition we discretise the lattice uniformly in $r$,
rather than $r^2$, which increases the effective resolution near the
symmetry axis. The new results reported used a maximum resolution of
256*1024 points in the $r, z$ directions.}).  In both dimensions they
behave similarly, and we find a maximum size localised black hole that
can `fit' in the circle dimension. The 6-d results provide evidence
that the non-uniform and black hole branches do indeed merge at a
topology changing solution.

\begin{figure}
\epsfig{file=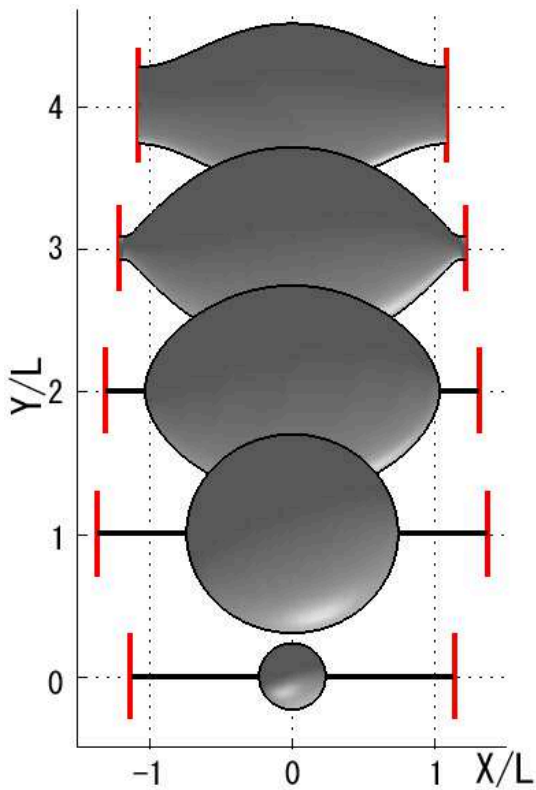,width=2.8in}
\caption{
Embeddings of the spatial horizon geometry of various 6-d BHs and NUSs
in 5-d Euclidean space (suitably projected onto the page). For BH
solutions we include the exposed symmetry axis in the embedding. The
red vertical lines are to be periodically identified, generating the
compact dimension.
\label{fig:embeddings}
}
\end{figure}

%
\section{Method}
\label{sec:method}
%

Both the non-uniform strings and black holes are static axisymmetric
geometries that can be written in the form,
\begin{equation}
ds^2 = - e^{2 A} dt^2 + e^{2 B} \left( dr^2 + dz^2 \right) + e^{2 C} r^2 d\Omega^2_{d-3}
\label{eq:metric}
\end{equation}
where $A,B,C$ are functions of $r,z$. We take these to vanish at large
radial coordinate $r$, and hence the geometry is asymptotically
$R^{1,d-2}\times S^1$. The circle coordinate $z$ has length $L$
asymptotically. We then employ a numerical method developed in
\cite{Wiseman:2001xt,Wiseman1,Kudoh:2003xz} which uses relaxation
techniques to solve for $A, B, C$, whilst ensuring all the Einstein
equations are satisfied. The reader is referred to \cite{Wiseman1} for
details of the procedure \footnote{See also \cite{Kunz} for work in
4-d utilising similar methods.}.

Following \cite{Townsend:2001rg,Kol:2003if,Harmark:2003eg,Harmark:2003dg},
because the solutions are not asymptotically flat, they are
characterised in terms of 2 asymptotic charges; the mass $M$ and a
dimensionless tension $n$, which then give a first law, $ d M = T d S + n M {d L}/{L} $ where $T, S$ are the black hole temperature and entropy.

In the following plots, we will fix the asymptotic circle length $L =
1$ for all solutions shown, and then use the quantity $n$ to
characterise the solutions. For small black holes $n \simeq 0$, and
for uniform strings $n = 1/(d-3)$.  For convenience we normalise the
thermodynamic quantities $T, S, M, n$ by their value for the critical
uniform string, $T_{crit}, S_{crit}, M_{crit}, n_{crit}$.

The most error prone part of the numerical calculation is extracting
the asymptotic form of the metric in order to compute $M,
n$. Particularly difficult is $n$ as it is constructed from the
difference of two relatively large quantities
\cite{Kudoh_Wiseman}. Here we compute both $n$ and $M$ using the first
law and Smarr formula from $T$ and $S$, which are conveniently
determined from the metric near the horizon.

%
\section{6-D Results}
\label{sec:result6d}
%

We now discuss the behaviour of 6-d localised black holes, and the
evidence that they join to the non-uniform branch. Firstly we consider
the geometry of the horizon.

\begin{figure}
\epsfig{file=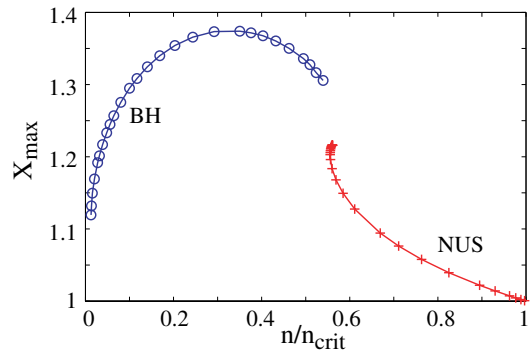,width=2.75in}
\caption{ 
Plot of $X_{max}$ for 6-d NUSs and BHs, consistent with merger
of the branches at $n/n_{crit} \simeq 0.55$.
\label{fig:geom1}
}
\end{figure}

We may embed the spatial horizon geometry as a surface of revolution
in 5-d Euclidean space. This is illustrated for several solutions in
figure \ref{fig:embeddings}. The intrinsic geometry of the embedded
surface is the same as that of the spatial sections of the 6-d
horizons. Note that for the black holes we also include the exposed
rotational symmetry axis in the embedding plots. The Euclidean
coordinate along the surface of revolution, $X$, should be thought of
as periodic with $-X_{max} < X < X_{max}$ being the fundamental domain
plotted. Then $X_{max}$ gives a coordinate invariant measure of the
`size' of the geometry near the symmetry axis, and we plot this in
figure~\ref{fig:geom1}.

\begin{figure}
\epsfig{file=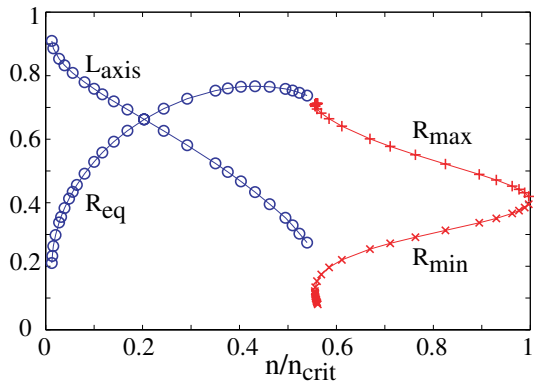,width=2.75in}
\caption{ 
Plot of horizon geometric quantities for 6-d solutions.  Branches are
consistent with a topology changing merger where both $L_{axis}$ and
$R_{min}$ go to zero, and $R_{eq}$ tends to $R_{max}$. All solutions
have $L = 1$.
\label{fig:geom2}
}
\end{figure}

For non-uniform solutions we may compute the maximum and minimum radii
of the horizon $R_{max,min}$. Analogously, for the black holes we have
$R_{eq}$, the equatorial radius of the horizon, and $L_{axis}$, the
proper length along the exposed symmetry axis. These are plotted in
figure~\ref{fig:geom2}.

From these graphs we see firstly that for increasing $n$ the
equatorial radius of the black holes reaches a maximum and then starts
to decrease, implying that there is indeed a maximum `size' localised
solution that can `fit'. Secondly, from the horizon geometry it is
plausible that the non-uniform and black hole branches merge around
$n/n_{crit} \simeq 0.55$, where the value of $R_{eq}$ appears to tend
to $R_{max}$, and $X_{max}$ 
is consistent with an extrapolation that agrees between the branches
for this value of $n/n_{crit}$.

\begin{figure}
\epsfig{file=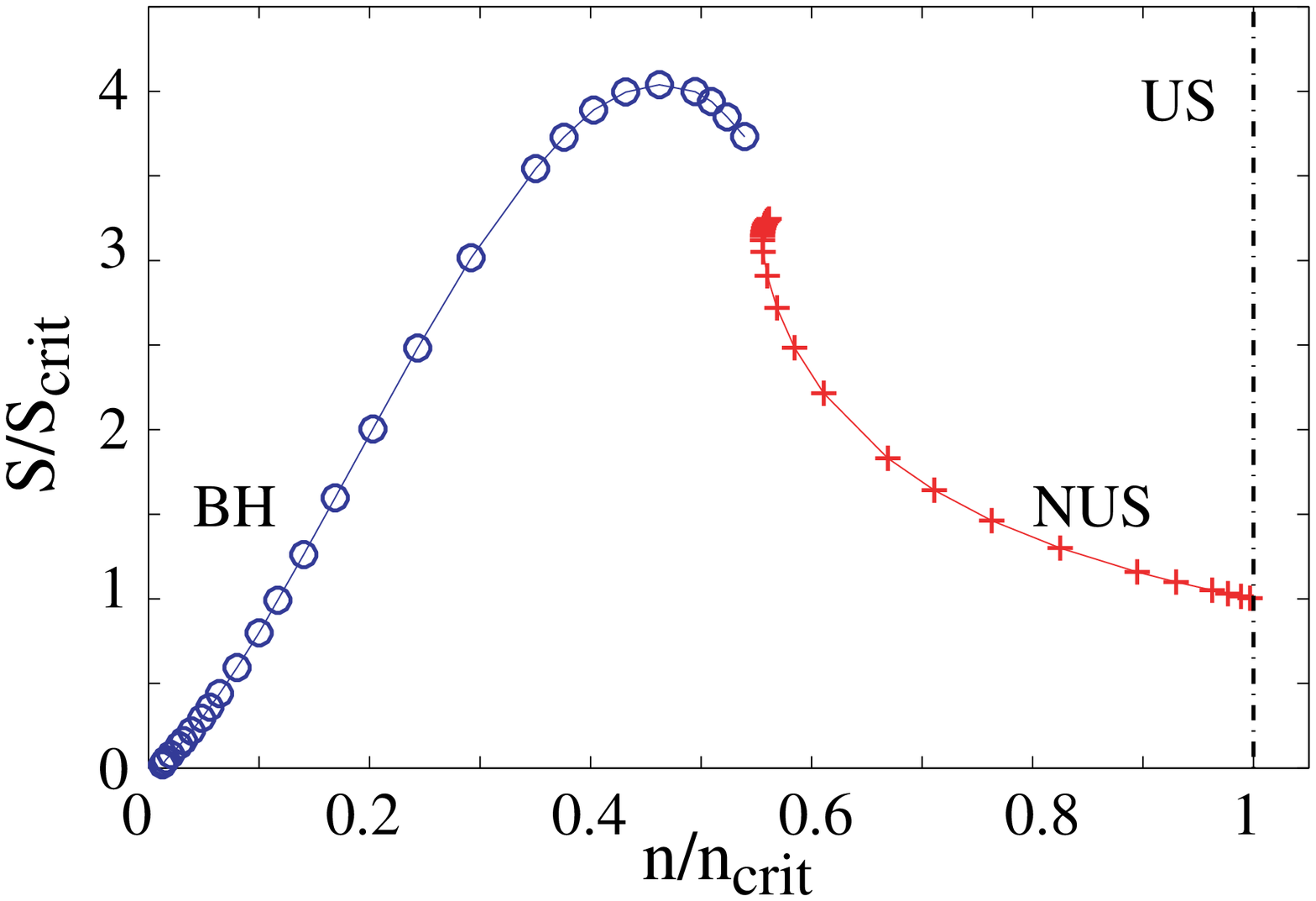,width=2.75in}
\epsfig{file=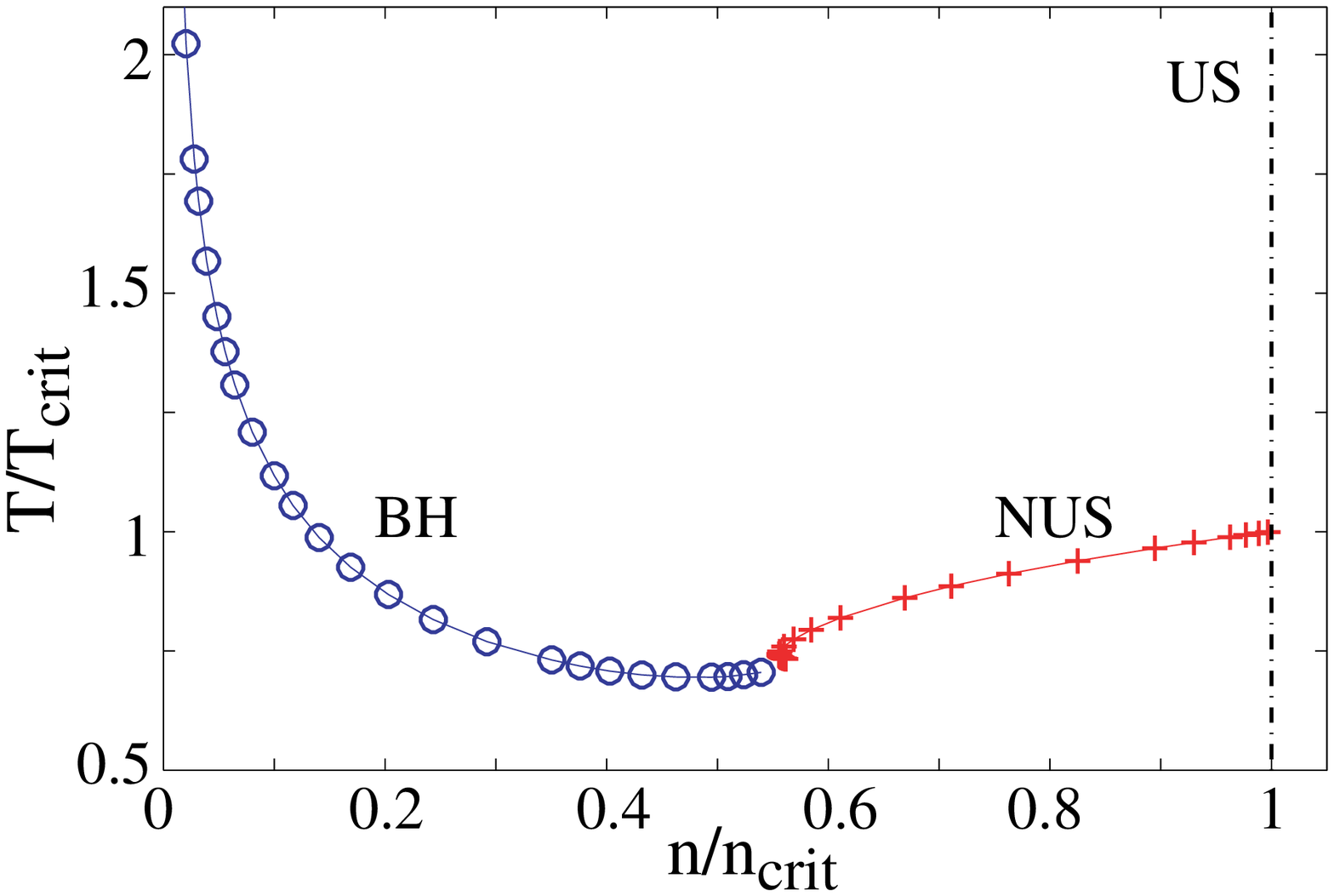,width=2.75in}
\caption{ 
Entropy and temperature for 6-d solutions.
\label{fig:T_and_S}
}
\end{figure}

\begin{figure}
\epsfig{file=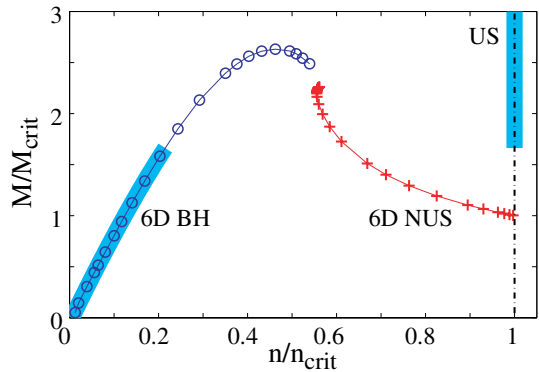,width=2.8in}
\caption{ 
Mass against $n$ for 6-d solutions. The highlighting indicates which
branch is entropically favoured for a given mass (see figure
\ref{fig:SM}).
\label{fig:mass6}
}
\end{figure}

Now we consider thermodynamic quantities. In figures \ref{fig:T_and_S}
and \ref{fig:mass6} we plot the entropy, temperature and mass of the
solutions against $n$. Mirroring the behaviour of $R_{eq}$, we see the
entropy and mass of the black holes reach a maximum and then decrease
with increasing $n$. From this thermodynamic data we again clearly see
evidence the branches merge.

Note that in our previous work \cite{Kudoh_Wiseman} constructing the
6-d black holes the solutions were only found in the regime where
$R_{eq}$ and the mass were increasing with $n$, and hence it was not
clear that the two branches could unify.

\begin{figure}
\epsfig{file=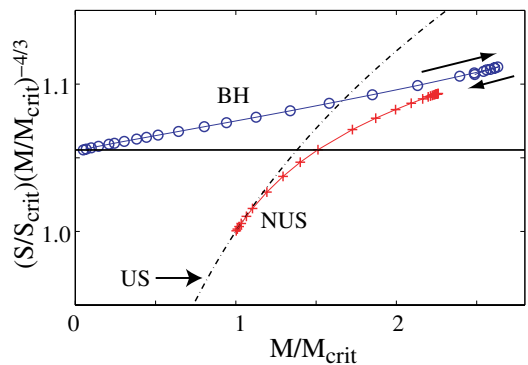,width=2.75in}
\caption{ 
Plot of $S M^{-4/3}$ against mass for the 6-d solutions. 6-d
Schwarzschild behaviour is $S \propto M^{4/3}$ so represents a
horizontal line. We see the localised BHs are entropically favoured
for a given mass, for $M < 1.7 M_{\mathrm{crit}}$. Above this mass the
uniform strings are favoured. The NUSs are never dominant.
\label{fig:SM}
}
\end{figure}

For very small masses the localised black holes are entropically
favoured, and for very large masses only the uniform strings exist. In
figure \ref{fig:SM} we plot the entropy against mass for the 3
branches. We see uniform strings become entropically favoured, for a
given mass, at masses above that of the critical uniform string, but
below that of the maximum mass localised black hole.

Thus we see that the branches appear consistent with merger at
$n/n_{crit} \simeq 0.55$. Let us now assume that this occurs via a
conical transition. We can then measure how far from the transition
point the solutions are by estimating the geometric resolution of the
cone. For the non-uniform strings this is given by the minimal radius
of the horizon, $R_{min}$, which for the most non-uniform string found
in \cite{Wiseman1} was $R_{min} \simeq 0.08$ in our units where $L =
1$. The resolution for the black hole is given by the proper distance
of exposed symmetry axis, $L_{axis}$. For these 6-d solutions the
smallest resolution found was $L_{axis} \simeq 0.27$. Therefore the
most non-uniform strings found are still considerably closer to the
assumed transition point than the most extreme localised black holes
found here. For these non-uniform solutions the emergence of the cone
geometry has been numerically demonstrated
\cite{Kol_Wiseman}. Repeating this for our new black hole solutions
does indeed show an increase in curvature on the axis consistent with
an emerging cone, but as the solutions are `further' from the
transition point this increase in curvature is still not large enough
to be clearly distinguished from the background curvature of the black
hole geometry.
Thus whilst our new 6-d black hole data is very suggestive the
solution branches merge, it still cannot confirm that the detailed
merger from the black hole side is via a conical transition. Indeed in
Kol's original picture \cite{Kol1} we note that the cone may act only
as an approximate local model for the merger, and the detailed
behaviour very close to the point where the horizon pinches off may
have a complicated behaviour that cannot necessarily be thought of as
being smoothly resolvable.

\section{5-D Results}
\label{sec:result5d}

\begin{figure}
\epsfig{file=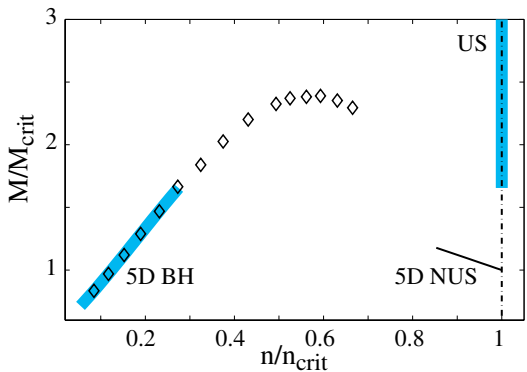,width=2.75in}
\caption{ 
Mass against $n$ for new 5-d localised BHs. Highlighting indicates
entropically favoured solution at a given mass.
\label{fig:mass5}
}
\end{figure}

We now briefly discuss the 5-d black hole solutions. All quantities
behave in an analogous manner to their counterparts for the 6-d
solutions. Here we have simply plotted the mass against $n$ in figure
\ref{fig:mass5}. Note the mass increases past the critical uniform
string mass with increasing $n$ (extending the previous numerical
solutions in 5d \cite{Sorkin:2003ka}), reaching a maximum and then
decreases, again presumably to join the non-uniform branch. \footnote{
We have been unable to extend the methods of \cite{Wiseman1} to
construct this 5-d non-uniform branch. The method works with the
asymptotic boundary at a relatively small radial coordinate location,
but this boundary cannot be removed to large coordinate radius without
an apparent numerical instability occurring - something that does not
occur in 6 or more dimensions.  }
\\

%
{\bf{Acknowledgements}}
We would like to thank Ofer Aharony, Gary Horowitz, Barak Kol, Jo
Marsano and Shiraz Minwalla for enjoyable and enlightening
discussions. TW was supported by the David and Lucille Packard
Foundation grant number 2000-13869A.  HK is supported by the JSPS.
Numerical computations were carried out at the YITP and NAO.

%
%


%
\end{document}